\documentclass[aip,amsmath,amssymb,reprint,nofootinbib,floatfix]{revtex4-1}

\usepackage{graphicx} 
\usepackage{dcolumn}
\usepackage{bm}

\usepackage[utf8]{inputenc}
\usepackage[T1]{fontenc}
\usepackage{mathptmx}
\usepackage{etoolbox}

\begin{document}

\title[Effects of Disorder on Electron Heating]%
{Effects of Disorder on Electron Heating in Ultracold Plasmas}

\author{Yurii V. Dumin}
\email[Corresponding author's electronic mail: \\ ]{dumin@pks.mpg.de,
dumin@yahoo.com}
\affiliation{HSE University, Faculty of Physics, \\
Myasnitskaya ul.\ 20, 101000 Moscow, Russia}
\affiliation{Space Research Institute of Russian Academy of Sciences, \\
Profsoyuznaya str.\ 84/32, 117997 Moscow, Russia}
\affiliation{Lomonosov Moscow State University,
Sternberg Astronomical Institute, \\
Universitetskii prosp.\ 13, 119234 Moscow, Russia}
\author{Anastasiia T. Lukashenko}
\email[Electronic mail: ]{lukashenko@dec1.sinp.msu.ru, a_lu@mail.ru}
\affiliation{Lomonosov Moscow State University,
Skobeltsyn Institute of Nuclear Physics, \\
Leninskie gory, GSP-1, 119991 Moscow, Russia}

\begin{abstract}
Starting from the beginning of their research in the early 2000's,
the ultracold plasmas were considered as a promising tool to achieve
considerable values of the Coulomb coupling parameter for electrons.
Unfortunately, this was found to be precluded by a sharp spontaneous increase
of temperature, which was often attributed to the so-called
disorder-induced heating (DIH).
It is the aim of the present paper to quantify the effect of spontaneous
heating as function of the initial ionic disorder and, thereby, to estimate
the efficiency of its mitigation, \textit{e.g.}, by the Rydberg blockade.
As a result of the performed simulations, we found that the dynamics of
electrons exhibited a well-expressed transition from the case of
the quasi-regular arrangement of ions to the disordered one;
the magnitude of the effect being about~30{\%}.
Thereby, we can conclude that the two-step formation of ultracold
plasmas---involving the intermediate stage of the blockaded Rydberg
gas---can really serve as a tool to increase the degree of Coulomb
coupling, but the efficiency of this method is moderate.
\end{abstract}


\maketitle

\section{Introduction}
\label{sec:Intro}

The so-called ultracold plasmas are neutral systems of charged particles
with typical electronic temperatures from a few to several tens of Kelvin,
which are obtained by a photoionization of gases cooled in the magneto-optical
traps; \textit{e.g.}, reviews~\cite{Gould01,Bergeson03,Killian07a,Killian07b}.
The experimental realization of such plasmas became feasible in the very
late 1990's and early 2000's, and they opened a new area of research in
the non-ideal plasma physics~\cite{Ichimaru82,Fortov00}.

It was initially expected that the extremely high values of the Coulomb
coupling parameter
\begin{equation}
\Gamma_{\! e} \approx \frac{\langle U \rangle}{\langle K \rangle}
\label{eq:Gamma_definition}
\end{equation}
(where $ K $ and $ U $ are the kinetic and potential energies of an electron)
can be achieved in this way.
Really, if energy of the ionizing laser irradiation was chosen to be slightly
above the ionization threshold of the cold neutral atoms, the initial kinetic
energy of the released electrons would be very low.
Therefore, it was expected in the first experiments~\cite{Killian99} that very
large values of the coupling parameter~(\ref{eq:Gamma_definition}) could be
obtained, \textit{e.g.}, tens or hundreds.
Unfortunately, it was quickly recognized that the situation is not so simple:
In fact, the cold photoelectrons are quickly accelerated by the electric fields
of nearby ions, and their temperature spontaneously increase by many times.

A simple pictorial explanation of this effect is the so-called disorder-induced
heating (DIH)~\cite{Gericke03}:
The charged particles (firstly, electrons and, at the longer time scale, also
the ions) tend to move to the positions with minimal potential energy.
As a result---since the total energy of the system should be conserved---%
kinetic energy of the particles will increase.
Therefore, DIH looks like an unavoidable effect, limiting the temperature
from below.

However, it was suggested by the same authors~\cite{Gericke03} that one can
get around the DIH effect by preparation of the system of charged particles
in the ``correlated'' state with a reduced Coulomb energy.
An attempt of realization of this approach was undertaken a decade later in
the experiment~\cite{Robert13}:
Namely, cold neutral atoms were initially transferred to the state of
Rydberg blockade, where the already excited atoms---due to their strong
electric fields---shift energy levels of the nearby atoms, thereby prohibiting
their excitation to the same Rydberg state~\cite{Lukin01,Tong04,Singer04}.
As a result, a quasi-regular arrangement of the excited atoms is formed,
where their close location to each other is excluded (\textit{e.g.},
Fig.~3 in paper~\cite{Dumin22}).
Next, after photoionization of Rydberg atoms, the resulting ions will
also be well separated from each other.
Therefore, one can expect that DIH will no longer take place, because the
system of ions is already in the ``quasi-crystalline'' state with minimal
potential energy.

\begin{figure*}
\includegraphics[width=0.95\textwidth]{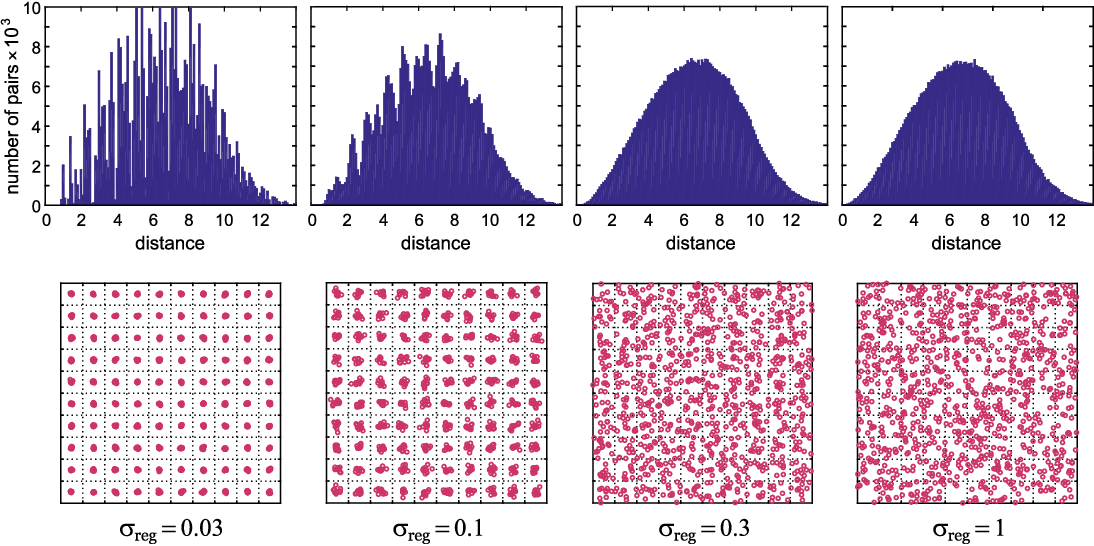}
\caption{\label{fig:Histogr_arrang}
Examples of the ionic arrangement in the $xy$-plane (bottom row) and
the corresponding histograms of the interparticle separation (top row)
for the quasi-regular distributions with different degrees of
disorder~$ {\sigma}_{\rm reg} $.
For convenience, a grid of dotted lines illustrates a characteristic
space per one particle in the perfect cubic lattice.}
\end{figure*}

The experiment~\cite{Robert13} was really able to trace a plasma formation
from the blockaded Rydberg gas, but it remained unclear if the resulting
electron temperature was appreciably reduced (and, respectively, the Coulomb
coupling parameter increased).
This was caused, firstly, by the fact that the ionization proceeded mostly
by the spontaneous avalanche process (and, therefore, did not strictly
preserve the initial quasi-crystalline arrangement of the Rydberg atoms) and,
secondly, by the insufficient diagnostic capabilities to measure the electron
temperature (S.~Whitlock, private communication).
Besides, there was no a clear theoretical prediction how much could be the
magnitude of the expected reduction in temperature.
Surprisingly, while the DIH mechanism was discussed for the first time about
20~years ago~\cite{Gericke03}, it was subsequently studied mostly for
ions~\cite{Pohl04,Pohl05,Donko09,Murphy15,Murphy16}, and there are no reliable
calculations of its influence on the electron temperature till now.

So, it is the aim of the present paper to quantify the corresponding effect in
electrons as function of the initial disorder.
In fact, this problem was partially touched in our previous work about the
clusterized plasmas~\cite{Dumin22}.
As a particular case of the nontrivial arrangement of the background ions,
we considered there also their quasi-regular distribution and found that
the resulting electron temperature was somewhat reduced.
However, the degree of such a reduction was comparable to the uncertainty
(r.m.s.\ variation) of that simulations.
Besides, it remained unclear how the corresponding effect depends on the degree
of disorder and, particularly, how it tends to the purely random case when
the disorder becomes sufficiently large.
In the present work---due to much better accuracy of the simulations---all
these issues will be resolved.

\section{Numerical Model}
\label{sec:Model}

A quite general model of ionic background with different degrees of disorder
can be formulated by the following way:
Let us consider initially a perfectly cubic lattice of ions.
The size of its cell~$ l $ will be used from here on as the unit of length,
and all the distances will be normalized accordingly.
Next, let each ion be shifted from its original position by a distance
given by the normal (Gaussian) distribution with r.m.s.\
deviation~$ {\sigma}_{\rm reg} $, as illustrated in the bottom row of
Fig.~\ref{fig:Histogr_arrang}.
Then, at $ {\sigma}_{\rm reg} \ll 1 $ (in dimensionless units) the ionic
distribution is quasi-regular; but it becomes more and more disordered when
$ {\sigma}_{\rm reg} $ increases; and finally, at
$ {\sigma}_{\rm reg} \sim 1 $, we evidently get a completely random
distribution.

These properties are well expressed in the histograms of interparticle
separation, shown in the top row of Fig.~\ref{fig:Histogr_arrang}.
For example, at $ {\sigma}_{\rm reg} \ll 1 $ one can see a series of sharp
peaks, which represent a set of the preferable interparticle distances in the
quasi-regular lattices.
Next, when the disorder increases, these peaks are gradually smoothed out;
and finally, at $ {\sigma}_{\rm reg} \sim 1 $, the histograms take
the Gaussian shape, which is typical for a random distribution.

To study dynamics of electrons against the above-mentioned kinds of
ionic background, we shall numerically integrate their equations of motion:
\begin{equation}
m_e \frac{d^2 {\bf r}_i}{dt^2} =
  - \sum_{j} e^2 \frac{ {\bf r}_i - {\bf R}_j }{ | {\bf r}_i - {\bf R}_j |^3 }
  + \sum_{k \neq i} e^2
    \frac{ {\bf r}_i - {\bf r}_k }{ | {\bf r}_i - {\bf r}_k |^3 } \, ,
\label{eq:Electron_motion}
\end{equation}
where
$ {\bf r}_i $ and $ {\bf R}_i $ are the electronic and ionic coordinates,
respectively; and $ m_e $ and $ e $~are the electron mass and charge.
The ions are assumed to be immobile, because we are interested only in the
sufficiently short time intervals.
The initial electron coordinates~$ {\bf r}_i(0) $ are given by the uniform
statistical distribution (\textit{i.e.}, the electrons are randomly
distributed with a uniform average density).
The initial electron velocities~$ {\bf v}_i(0) $ are specified by the normal
(Maxwellian) law with the r.m.s.\ variation~$ {\sigma}_{\! v} $:
\begin{equation}
f (v_{\alpha}) = \frac{1}{\sqrt{2 \pi} {\sigma}_v}
  \exp \bigg[ \! -\frac{v_{\alpha}^2}{2 {\sigma}_v^2} \bigg] \, ,
\label{eq:sigma_v_Def}
\end{equation}
where
$ \alpha $~is $ x $, $ y $, or~$ z $.
In the ideal plasmas, the above-mentioned parameter is evidently proportional to
the square root of electron temperature.
However, the situation can be more tricky in the non-ideal case.
This is the reason why we prefer to speak here just about the r.m.s.\
variation.

Strictly speaking, the initial values of the electron coordinates and
velocities strongly depend on the details of the ionization process.
For example, in the case of instantaneous photoionization, the initial
electron positions will be strongly correlated with the positions of
ions~\cite{Niffenegger11}, while distribution of the electron velocities
will look like the delta-function.
However, if the photoionization takes some time, the released electrons
should be somewhat mixed between the ions and the distribution of their
velocities smoothed out, resulting in the Maxwellian form.

Next, we shall use the perfectly reflective boundary conditions, and
the Coulomb interactions will be calculated within the fixed simulation box.
An alternative choice---used in the most of the previous works on
ultracold plasmas---is to employ the periodic boundary conditions, when
a particle leaving the simulation box through one of its boundaries
simultaneously enters it through the opposite boundary.
The Coulomb interactions are usually calculated in such a case by
the ``wrapping'' algorithm~\cite{Niffenegger11}, when each particle interacts
with other particles within a moving cube of size~$ \pm L/2 $, centered at
that particle (where $ L $~is the total size of the simulation box).

In principle, none of these options is perfect:
The reflective boundaries evidently affect bulk properties of the plasmas.
However, as was shown in our previous work~\cite{Dumin22}, averaging over
a sufficient number of initial conditions substantially mitigates this
problem.
On the other hand, the periodic boundary conditions with the above-mentioned
wrapping procedure should not distort the bulk properties.
Unfortunately, a closer inspection shows that swapping of a charged particle
between the boundaries of a moving box results in the abrupt unphysical
change in the direction of Coulomb force between the respective pair of
particles.
From this point of view, the reflective boundaries look better because their
effect on the bulk properties has a clear physical meaning~\cite{Mayorov94},
and it will evidently disappear with increasing the number of particles.

Besides, in the context of our simulations, a decisive advantage of the
reflective boundary conditions is that they are well consistent with the
algorithms of numerical integration with the adaptive stepsize control (ASSC),
such as the subroutines \texttt{odeint}, \texttt{rkck}, and \texttt{rkqs} from
the Numerical Recipes~\cite{Press92}.
These subroutines were already used in our previous work~\cite{Dumin22} and
demonstrated a perfect performance.
Particularly, they are able to work with the ``true'' (singular) Coulomb
potentials, without a need for their cut-off or ``softening'' at small
distances.
This excludes any artifacts caused by the modified potentials.
(Let us mention that dealing with singular potentials without the ASSC
algorithms requires the huge computational resources~\cite{Dumin11}.)
Unfortunately, ASSC becomes unreliable in the case of periodic boundary
conditions because of the above-mentioned abrupt jumps of the Coulomb forces.

All the results will be presented below in dimensionless units, introduced
by the following way:
a unit of length is a mean distance between the ions~$ l $
(the corresponding unitary cells are marked by dotted lines in
Fig.~\ref{fig:Histogr_arrang});
a unit of time, up to numerical factor on the order of unity, is
the inverse plasma frequency,
\begin{equation}
\tau = \sqrt{\frac{m_e l^3}{e^2}} =
  \frac{\sqrt{4 \pi}}{{\omega}_{\rm pl}} \, ;
\label{eq:Def_tau}
\end{equation}
and a unit of energy is the characteristic Coulomb energy at the
interparticle distance, $ U = e^2 / l $.
The dimensionless quantities expressed in these units will be denoted by hats.

Let us emphasize that the unit of time~(\ref{eq:Def_tau}) has a deep physical
meaning:
If the initial plasma state is substantially overcooled, \textit{i.e.},
the electron velocities are sufficiently small, then each electron will
fall onto the nearest ion.
The corresponding Keplerian trajectory will be strongly elliptical, and
the ion will be located at the focus of this ellipse opposite to the initial
position of the electron.
Next, it should be taken into account that a period of revolution along
the Keplerian trajectory depends only on its major axis~$ a $ and
equals~\cite{Landau76}:
$ T_{\rm Kep} = (\pi / \sqrt{2}) \, m_e^{1/2} a^{3/2} \! / e $.
A typical distance from the electron to the nearest ion should be
about~$ l / 2 $.
Then, taking this quantity as the major axis, we get:
$ T_{\rm Kep} = (\pi / 4) \tau $, where the numerical factor is very
close to unity.
Therefore, a characteristic fall time of any electron should be about
one half of the unit of time given by formula~(\ref{eq:Def_tau}).
In this sense, the accelerated electrons behave ``synchronously''
and---up to numerical factor about unity---are characterized by
the inverse plasma (Langmuir) frequency.
(This argumentation evidently refers only to the electron dynamics.
Treatment of ions, especially when they experience the Debye screening,
requires a more complex analysis, which was undertaken, for example, in
paper~\cite{Bergeson11}.)
Besides, the above-mentioned ``synchronous motion'' does not imply
the applicability of the mean-field approximation, which can be used only
for the sufficiently ideal and equilibrium  plasmas.

Finally, the electron temperature is assumed to be related to the kinetic
energy per electron by the same formula as for an ideal gas,
$ T_e = (2/3)\,K / k_{\rm B} $.
Although we deal here with a substantially non-ideal case, it was found in
paper~\cite{Niffenegger11} that such definition reasonably agrees with a more
elaborated derivation of~$ T_e $, based on the approximation of the simulated
velocity distributions by the Maxwellian ones.
A theoretical explanation of this fact can be found in paper~\cite{Dumin00}.

\section{Results of the Simulations}
\label{sec:Results}

Our simulations were performed for a system of 1000~particles of each kind
(electrons and ions), \textit{i.e.}, the simulated volume was composed of
$ 10 \times 10 \times 10 $~unitary cells, as depicted in
Fig.~\ref{fig:Histogr_arrang}.
This is 8~times greater than in the previous simulations of the clusterized
plasmas~\cite{Dumin22}: since influence of the quasi-regular ionic arrangement
is a finer effect, we had to increase the number of simulated particles.

In the particular case presented below, we used $ {\hat{\sigma}}_{\! v} = 0.3 $,
which implies that the initial kinetic energy of electrons was about an order
of magnitude less than their potential (Coulomb) energy, \textit{i.e.},
the plasma was substantially overcooled.
It should be expected that final results will be insensitive to the particular
degree of overcooling, because the major effect comes from the subsequent
spontaneous heating.
(Dependence of the simulations on~$ {\hat{\sigma}}_{\! v} $ is discussed
in more detail in Appendix~\ref{sec:Init_temp}.)
In other words, the plasma was already in the non-ideal (strongly-coupled)
state, and our aim was to check how this state will survive against the
subsequent heating.

The simulations were performed for the following degrees of disorder:
$ {\sigma}_{\rm reg} = 0.01, 0.03, 0.1, 0,3, 1 $ and 3, as well as for
the purely random ionic distribution.
To get the statistically significant results, five versions of initial
conditions---randomly generated both for the ions and electrons---were used
for each of these values.
Despite dealing with singular interparticle potentials, the algorithm of
the adaptive stepsize control enabled us to get a sufficiently high accuracy
of integration.
It was estimated, as usual, by a conservation of the total energy of
the system.
In the worst case, such error was 1.4{\%}, but usually by one or two orders
of magnitude better.

\begin{figure}
\includegraphics[width=0.8\columnwidth]{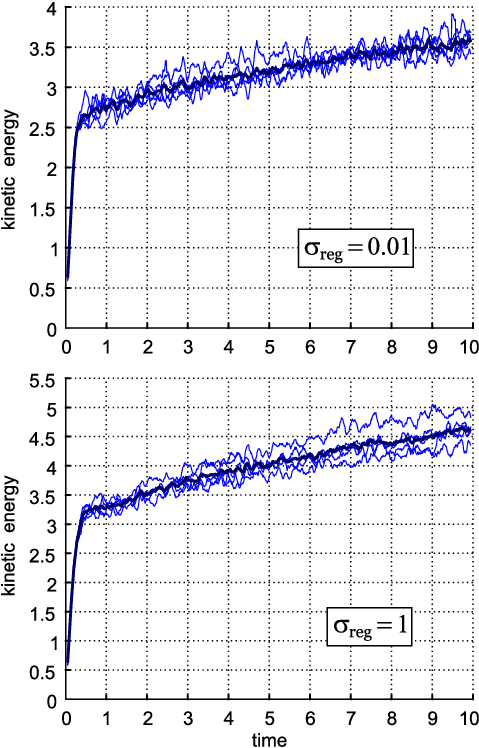}
\caption{\label{fig:Ener_time_indiv}
Individual profiles of the electron kinetic energy as function of time
(thin curves) and their average behavior (thick curves) at
$ {\sigma}_{\rm reg} = 0.01 $ and 1 (top and bottom panels, respectively).}
\end{figure}

Examples of the simulated temporal behavior of the dimensionless kinetic
energy of electrons are presented in Fig.~\ref{fig:Ener_time_indiv}.
To avoid obscuring the plots with a lot of sharp peaks, caused by
the close interparticle collisions, they were smoothed out over a running
window of width $ {\Delta}\hat{t} = 0.1 $.
In principle, a general shape of these curves---a quick initial jump
at the timescale $ \hat{t} \approx 0.5 $ followed by a much longer gradual
increase---was well known already from the pioneering works by Kuzmin and
O'Neil~\cite{Kuzmin02a,Kuzmin02b}.
It is the aim of our study to reveal how they depend on the arrangement of
the ionic background.

\begin{figure}
\includegraphics[width=0.85\columnwidth]{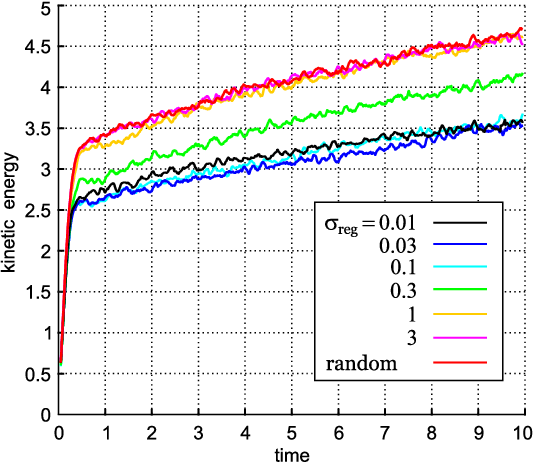}
\caption{\label{fig:Ener_time_avr}
Average (over 5~realizations) temporal profiles of the electron kinetic
energy~$ \langle \hat{K} \rangle $ for the entire set of the disorder
parameters~$ {\sigma}_{\rm reg} $.}
\end{figure}

Figure~\ref{fig:Ener_time_avr} shows the average temporal profiles of
the electron kinetic energy for the entire variety of the disorder
parameters~$ {\sigma}_{\rm reg} $, ranging from an almost perfect
ionic lattice to the completely random distribution.
One can see here three types of the curves:
Type~I corresponds to the small values of~$ {\sigma}_{\rm reg} $
(0.01, 0.03, and 0.1), \textit{i.e.}, the weakly distorted lattices.
Type~II with $ {\sigma}_{\rm reg} = 0.3 $ is the intermediate case,
corresponding to the moderately distorted lattice.
At last, the curves of Type~III belong either to the cases of strongly
distorted lattices, $ {\sigma}_{\rm reg} = 1 $ and 3,
or to a completely random distribution.
As could be naturally expected, Type~II indicates a noticeable change
in the temporal behavior of the electron kinetic
energy~$ \langle \hat{K} \rangle $.
Namely, when $ {\sigma}_{\rm reg} $ increases, the initial jump
(at the timescale $ 0 \leq \hat{t} \lesssim 0.5 $) becomes
more pronounced.
The subsequent gradual increase in~$ \langle \hat{K} \rangle $
at $ \hat{t} \gtrsim 0.5 $, in principle, also changes but insignificantly.

Referring to the histograms of interparticle separation, presented in the top
row of Fig.~\ref{fig:Histogr_arrang}, one can see that at small values
of~$ {\sigma}_{\rm reg} $ (\textit{e.g.}, 0.03 and 0.1) they exhibit
a series of sharp peaks, which are typical for the crystalline-like structures.
On the other hand, at the large values of~$ {\sigma}_{\rm reg} $
(\textit{e.g.}, 1) their shape closely resembles the purely-random Gaussian
distribution.
It is interesting that in the intermediate case $ {\sigma}_{\rm reg} = 0.3 $
the histogram is almost Gaussian, but
the curve $ \langle \hat{K} \rangle (\hat{t}) $ in Fig.~\ref{fig:Ener_time_avr}
exhibits a clearly distinct behavior.
In fact, a closer inspection of the patterns of ionic arrangement in the bottom
row of Fig.~\ref{fig:Histogr_arrang} shows that the case
$ {\sigma}_{\mbox{reg}} = 0.3 $ preserves some features of the quasi-regularity:
namely, there are no occasional ``voids'', typical for the purely random
distributions.

\begin{figure}
\includegraphics[width=0.85\columnwidth]{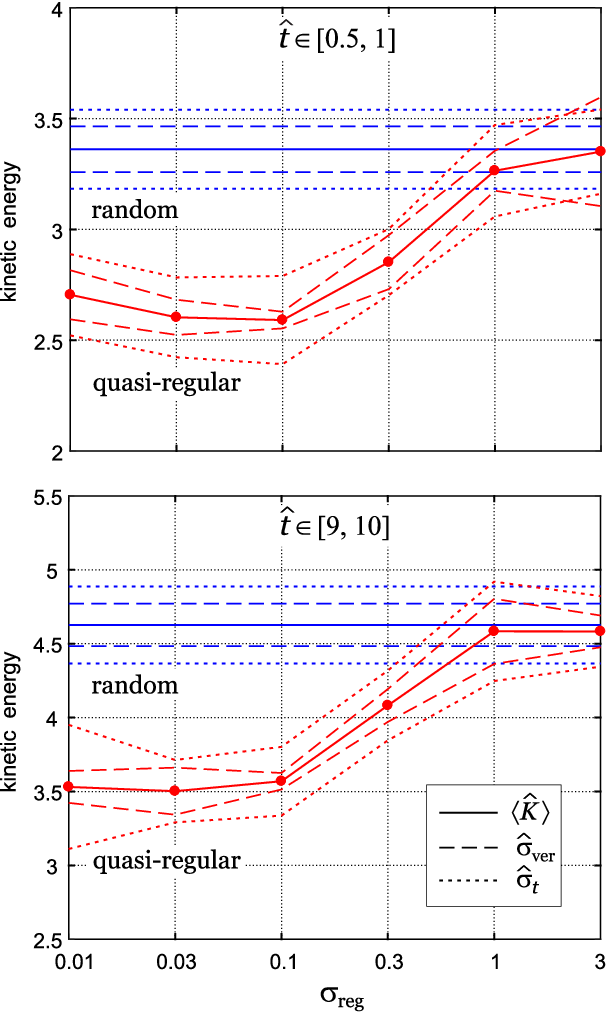}
\caption{\label{fig:Avr_ener_sigma}
Average values of the electron kinetic energy established at the time intervals
[0.5, 1] (top panel) and [9, 10] (bottom panel), as well as their r.m.s.\
deviations, as functions of the disorder parameter~$ {\sigma}_{\mbox{reg}} $.}
\end{figure}

To quantify the three above-mentioned types of the temporal
behavior~$ \langle \hat{K} \rangle (\hat{t}) $, we calculated average values
of the electron kinetic  energy at two time intervals:
$ \hat{t} \in [0.5, 1] $ and $ \hat{t} \in [9, 10] $.
The first of these quantities characterizes a magnitude of the initial jump
of~$ \hat{K} $, which was often attributed just to the disorder-induced
heating; while the second quantity represents the total increase
in~$ \hat{K} $ after a sufficiently long period of evolution,
which includes also a subsequent heating of plasma due to recombination.
First of all, this is the three-body recombination, which was discussed by
various authors starting from the first experiments with ultracold plasmas.
However, as follows from our recent simulations of the clusterized
plasmas~\cite{Dumin22}, the multi-body processes (involving two electrons and
more than one ion) should be also important in the non-ideal case:
Really, since the rate of recombination substantially depended on the degree
of clusterization, a simultaneous scattering of two electrons by two or more
ions came into play there.
The above-mentioned quantities are presented in Fig.~\ref{fig:Avr_ener_sigma}
as functions of~$ {\sigma}_{\mbox{reg}} $.

To estimate a statistical significance of the simulations, we plotted in
this figure also the r.m.s.\ variations of the average energy with respect
to various versions of the initial conditions~$ \hat{{\sigma}}_{\mbox{ver}} $
and with respect to time~$ \hat{{\sigma}}_{t} $ over the corresponding
intervals (for mathematical details, see Appendix~\ref{sec:Avr_quant}).
These quantities are shown by the dashed and dotted lines, respectively.
As is seen, $ {\hat{\sigma}}_{\mbox{ver}} $~is appreciably smaller
than~$ \hat{{\sigma}}_{t} $.
In other words, five versions of the initial conditions are quite sufficient
for the reliable averaging.
On the other hand, the temporal r.m.s.\ variation~$ \hat{{\sigma}}_{t} $ is
evidently unavoidable, and its effect can be reduced only by increasing
the number of particles in the simulated system.

\section{Discussion and Conclusions}
\label{sec:Discussion}

1. The main result of our study is the identification of the clear transition
of the electron dynamics from the case of quasi-regular ionic background
(\textit{e.g.}, caused by the Rydberg blockade, as discussed in the
Introduction) to the random one.
Let us mention that some reduction of the electron temperature in the
regularized ionic arrangement was found already in our previous
work~\cite{Dumin22}, but this effect in~$ \hat{K}(\hat{t}) $ was comparable
to the uncertainties~$ {\hat{\sigma}}_{\mbox{ver}} $
and~$ \hat{{\sigma}}_{t} $; see Table~1 and Fig.~8 in that paper.
Besides, it was rather surprising why there was no a well-expressed transition
of~$ \hat{K}(\hat{t}) $ to the purely random case
when~$ {\sigma}_{\mbox{reg}} $ tended to unity.
In the present study---due to the enhanced accuracy of simulations---%
this puzzle was resolved, and the clear transition was identified.

Unfortunately, the effect of reduction of the electron temperature is not
so large: as is seen in Fig.~\ref{fig:Avr_ener_sigma}, it is
approximately~30{\%} both immediately after the sharp jump, occurring at the
timescale of one half of the dimensionless unit, and at the longer time
interval, when a heat release due to recombination came into play.

It is interesting to mention that yet another method to reduce the electron
temperature (and, thereby, to increase the Coulomb coupling parameter)
was suggested in papers~\cite{Vanhaecke05,Crockett18}.
This is adding the Rydberg atoms with binding energies
$ | E_{\rm b} | \lesssim (2-3)\,k_{\rm B}T_e $ into the ultracold plasmas.
As a result, their inelastic collisions with free electrons will lead to
further excitation of the atoms and cooling of the electrons.
It was found in the above-cited papers that the overall efficiency of such
a process should be about~20{--}30{\%}, \textit{i.e.}, actually the same
as in the method based on the Rydberg blockade~\cite{Robert13}.

At last, reduction of the electron heating rate by a factor of~3 was obtained
by simulating the strongly-magnetized ultracold plasmas~\cite{Tiwari18}, when
the electron motion was effectively constrained to a single dimension.
However, since such plasmas involve a strong and long-lasting temperature
anisotropy, they represent substantially different physical systems as
compared to the ones considered in the present paper.

2. Yet another unexpected finding in  our simulations is that there was a rather
strong spontaneous heating of the ultracold plasma even in the case of almost
regular lattice, \textit{i.e.}, when there was essentially no disorder.
So, we believe that the concept of disorder-induced heating (DIH) may have
a limited scope of applicability.
An alternative explanation can be based, for example, on the concept of
``virialization''~\cite{Dumin00}, which was suggested even before DIH but
did not attract attention till now.
In that case, a sharp increase in temperature is attributed to the
establishment of virial distribution between the kinetic and potential
energies and, therefore, the rate of initial heating should not depend
appreciably on the degree of disorder.
Of course, a more detailed analysis of the behavior of different kinds of
energy should be performed to discriminate more definitively between
the concepts of DIH and virialization; this is planned to be done in
a separate paper.

3. A rather restrictive assumption in our simulations was imposition of
the reflective boundary conditions, which was necessary to ensure a smooth
operation of the numerical integration algorithms with the adaptive
stepsize control (ASSC).
In principle, the reflective walls might affect the bulk properties of
the simulated plasmas (especially, when the number of particles in the model
is not so large, and a considerable fraction of them is located near
the walls).
However, as was shown in our previous paper~\cite{Dumin22}, this problem can be
substantially mitigated by averaging over a sufficiently large set of initial
conditions.
Then, the results become almost independent on the number of particles in
the simulation cell.

Of course, it would be desirable to use in future simulations the periodic
boundary conditions, which are more relevant from the physical point of view.
A promising approach to reconcile the periodic boundary conditions
with ASSC might be the so-called Ewald summation of the Coulomb
interactions~\cite{Ziman72,Rapaport04}.
Then, there should be no sharp jumps of forces when a particle is transferred
from one boundary of the cell to another, and ASSC should work much better.
But the implementation of such approach evidently requires a lot of additional
work.

\begin{acknowledgments}

YVD is grateful to J.-M.~Rost and S.~Whitlock for stimulating this study,
as well as to A.A.~Bobrov, S.A.~Mayorov, P.R.~Levashov, U.~Saalmann, and
V.S.~Vorob'ev for fruitful discussions and advices.
We are also grateful to all anonymous referees for the careful analysis of
our results and a lot of valuable suggestions.

\end{acknowledgments}

\section*{Author Declarations}

\subsection*{Conflict of Interest}

The authors have no conflicts to disclose.

\subsection*{Author Contributions}

YVD suggested the theoretical concept, developed the corresponding
software, and prepared the manuscript; both authors performed the
simulations and analyzed the obtained results.

\section*{Data Availability}

The data that support the findings of this study are available from the
corresponding author upon reasonable request.

\appendix

\section{Dependence on the Initial Electron Temperature}
\label{sec:Init_temp}

All the results presented in the main part of this paper were obtained for
the case of sufficiently small initial dispersion of the electron velocities,
$ {\hat{\sigma}}_{\! v} = 0.3 $, since we were interested in the evolution of
substantially overcooled plasmas with significant values of the electron's
Coulomb coupling.
In general, it might be expected that such evolution will be almost independent
of the degree of overcooling as long as the initial kinetic energy remains
small in comparison with the potential one.
However, it is important to check this conjection.

\begin{figure}
\includegraphics[width=0.85\columnwidth]{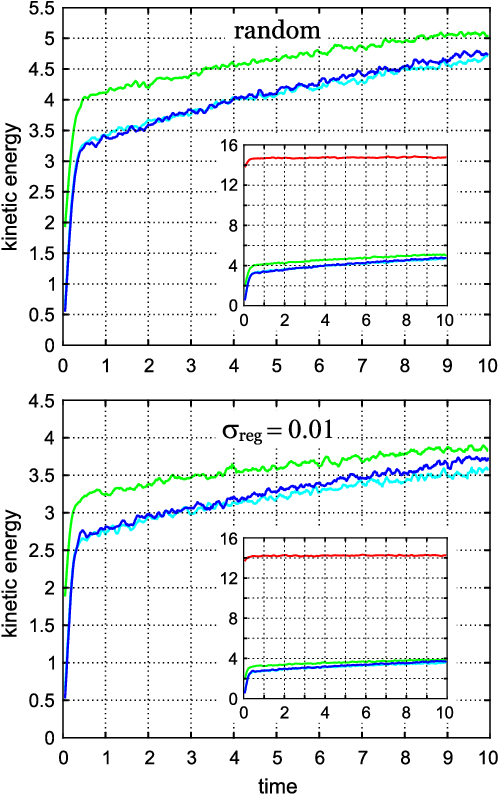}
\caption{\label{fig:Depend_init_temp}
Average (over 5~realizations) temporal behavior of the electron kinetic
energy~$ \langle \hat{K} \rangle $ at the initial dispersion of the
electron velocities $ {\hat{\sigma}}_{\! v} = 0.1 $ (blue curves),
0.3 (cyan), 1.0 (green) and 3.0 (red) for the purely random
spatial distribution of ions (top panel) and their almost regular
arrangement with $ {\sigma}_{\mbox{reg}} = 0.01 $ (bottom panel).}
\end{figure}

With this aim in view, we performed a set of additional simulations with
various values of initial dispersion:
$ {\hat{\sigma}}_{\! v} = 0.1, 0.3, 1.0 $, and 3.0, corresponding to the
initial electron kinetic energies about 0.01, 0.1, 1, and 10 (in the
dimensionless units, normalized to the characteristic potential energy).
The respective results are presented in Fig.~\ref{fig:Depend_init_temp}
for the limiting cases of a random ionic background (top panel) and
an almost unperturbed lattice with $ {\sigma}_{\mbox{reg}} = 0.01 $
(bottom panel).
The regimes of temperature evolution starting from the above-mentioned values
of~$ {\hat{\sigma}}_{\! v} $ are shown by blue, cyan, green, and red curves,
respectively.

As could be expected, in the case of strongly overcooled initial state
($ {\hat{\sigma}}_{\! v} \ll 1 $), a subsequent evolution of the electron
kinetic energy is almost independent of~$ {\hat{\sigma}}_{\! v} $:
Really, as is seen in both panels, the blue ($ {\hat{\sigma}}_{\! v} = 0.1 $)
and cyan ($ {\hat{\sigma}}_{\! v} = 0.3 $) curves are almost indistinguishable
from each other.
Moreover, the cyan curve is often invisible at all, since it is obscured by
the blue one.

However, when the initial plasma state was taken to be moderately coupled
(non-ideal), \textit{e.g.} $ {\hat{\sigma}}_{\! v} = 1.0 $,
there was a noticeable change in the temperature evolution:
it is seen that a green curve lies above the blue and cyan ones.
This is not surprising because the electrons possessed an additional energy
already in the initial state.
Besides, despite a sharper initial jump, a subsequent increase of
the temperature proceeds a bit slower.
This is also rather natural, because it is well-known that the rate of
three-body recombination considerably drops with increase in the electron
temperature~\cite{Massey56} (and the same should be expected for
the multi-body recombination).

At last, when the plasma was initially almost ideal (for example,
$ {\hat{\sigma}}_{\! v} = 3.0 $, \textit{i.e.}, its kinetic energy exceeded
the potential one by an order of magnitude), the corresponding red curve
lies well above other curves.
Moreover, it is almost horizontal, \textit{i.e.}, a heating due to
recombination is negligible.
This case is evidently irrelevant to the situation discussed in the main body
of the paper.

In principle, the fact that dynamics of the overcooled plasmas is almost
independent of the particular values of initial temperature was mentioned
already in paper~\cite{Tiwari18}.
However, that simulations were performed with the ``soft-core'' potentials
(\textit{i.e.}, Coulomb interactions between the electrons and ions were
artificially cut off at small distances).
Since the dynamics of overcooled electrons involves a lot of very close
encounters with ions, it is unclear in advance if the above-mentioned
independence will survive in the case of more realistic Coulomb interactions.
So, as follows from the present simulations, this really takes place.

\section{Calculation of the Averaged Quantities}
\label{sec:Avr_quant}

Since our article involved a number of various averaged quantities, it is
reasonable to discuss the particular formulas for their calculation.
For the sake of brevity, we shall omit here the hats denoting dimensionless
quantities.

Let $ K_i^{\rm (raw)} (t) $ be the original (``raw'') time series of
the simulated electron energies at the $ i $'th version of initial conditions.
As was already mentioned before, this quantity involves a lot of sharp peaks
caused by the close interparticle collisions, which are irrelevant to the
collective plasma behavior.
So, the first step is to perform their averaging over a running window of
width~$ \Delta \, t $,
\begin{equation}
K_i (t) = \frac{1}{\Delta \, t}
  \int\limits_{t - \Delta \, t / 2}^{t + \Delta \, t / 2}
  K_i^{\rm (raw)} (t') \, \mbox{d}\,t' \: .
\label{eq:K_i_t}
\end{equation}
Such averaged quantities~$ K_i (t) $ are shown, for example, by thin blue
curves in Fig.~\ref{fig:Ener_time_indiv}.

Besides, we introduce the electron kinetic energy averaged over various
versions of the initial conditions ($ i = 1, \dots , N $):
\begin{equation}
\langle K (t) \rangle =
  \frac{1}{N} \, \sum\limits_{i = 1}^{N} K_i (t) \: .
\label{eq:K_t}
\end{equation}
These quantities are shown by thick dark curves in
Fig.~\ref{fig:Ener_time_indiv} as well as by the thick colored curves in
Figs.~\ref{fig:Ener_time_avr} and~\ref{fig:Depend_init_temp}.

Next, to characterize the electron kinetic energy established after some
period of evolution starting from the $ i $'th version of initial conditions,
we perform its averaging over a specified time interval $ [t_1, t_2] $:
\begin{equation}
\overline{K_i} =
  \frac{1}{(t_2 - t_1)}
  \int\limits_{t_1}^{t_2} K_i^{\rm (raw)} (t') \, \mbox{d}\,t'
\: ;
\label{eq:K_i_timeint}
\end{equation}
and similarly we find the corresponding r.m.s.\ variation at the same time
interval:
\begin{equation}
\overline{{\sigma}_i} =
  \Bigg\{ \frac{1}{(t_2 - t_1)}
  \int\limits_{t_1}^{t_2} \big[ K_i^{\rm (raw)} (t') - \overline{K_i} \big]^2
  \, \mbox{d}\,t'
  \Bigg\}^{1/2} .
\label{eq:sigma_i_timeint}
\end{equation}
At last, the established energies are averaged over the initial conditions:
\begin{equation}
\langle \overline{K} \rangle =
  \frac{1}{N} \, \sum\limits_{i = 1}^{N} \overline{K_i} \: .
\label{eq:K_timeint}
\end{equation}
The respective quantities are shown by the solid lines and circles in
Fig.~\ref{fig:Avr_ener_sigma}.

Next, to characterize ``stability'' of the obtained mean values, we
introduce two kinds of their r.m.s.\ deviations, which are calculated by
the following way.
Firstly, a variation with respect to different versions of the initial
conditions is evidently defined as
\begin{equation}
{\sigma}_{\rm ver} =
  \bigg\{ \frac{1}{N - 1}
  \sum\limits_{i = 1}^{N}
  \Big[ \overline{K_i} - \langle \overline{K} \rangle \Big]^2
  \bigg\}^{1/2} .
\label{eq:sigma_ver}
\end{equation}
These quantities are shown in Fig.~\ref{fig:Avr_ener_sigma} by the red and blue
dashed lines.

The second important characteristics is a r.m.s.\ variation with respect to
time at the specified interval.
It is obtained by averaging the corresponding
quantities~(\ref{eq:sigma_i_timeint}) over the all initial conditions:
\begin{equation}
{\sigma}_{t} = 
  \frac{1}{N} \, \sum\limits_{i = 1}^{N} \overline{{\sigma}_i} \: .
\label{eq:sigma_t}
\end{equation}
These variations are plotted in Fig.~\ref{fig:Avr_ener_sigma} by the red and
blue dotted lines.

\subsection*{References:}
%

%

\end{document}